\documentclass[seceq,supplement]{ptptex}
\usepackage{epsfig}
\usepackage{graphicx}
\usepackage{latexsym}
\usepackage{wrapft}

\title{Lattice QCD at finite density}
\subtitle{A primer}    
\author{Maria-Paola \textsc{Lombardo}}

\inst{Istituto Nazionale di Fisica Nucleare, Italy}
\abst{
QCD  at  finite  density  presents  specific challenges  to  lattice  gauge
theory. Nonetheless,  a region  of the QCD  phase diagram up  to moderately
large  baryon chemical  potentials has  been successfully  explored  on the
lattice and new  results and idea are continuously  emerging.

I  will  outline  the  lattice  formulation  of  QCD,  introduce  the
calculational schemes currently  used to treat a nonzero  baryon density,
and  mention  lattice  methods   alternative  to
MonteCarlo,  including  the strong  coupling  expansion   which might  give
access to the the superconducting phase of QCD.  The results
for the critical line, and the different phases will be discussed
highlighting the strength of the different methods, as well as
the possible comparisons with  phenomenological models.}

\begin{document}
\maketitle

\section{Introduction}
Lattice discretization combined with importance sampling
affords the possibility of doing 
first principles calculations of the properties of
strongly interacting  matter \cite{intro}.
Questions to be addressed obviously include 
the study of QCD  with physical values of the quark masses,
and in the thermodynamic region within the range of current, and planned,
 experiments at BNL and CERN.

While addressing these phenomenological points, one is
lent to consider more theoretical questions --
patterns of chiral symmetry, mechanisms of confinement, topological
structures, gauge field dynamics...--
on  general grounds. To this end, it is not only legitimate,
but also useful (and sometimes mandatory),  
to consider QCD  in a larger parameter
space. This can be achieved  in a numerical simulation, where
not only we can tune experimental  parameters
such as baryon chemical potential, temperature, isospin chemical potential,
but we can also play with the number of colors and flavors, the value
of the bare masses, the gauge coupling,  
or we can make imaginary some external fields. In the following QCD will
thus be a generic name for any of these variants -- to be
specified whenever needed --  of the ``real world'' 
theory of strong interactions.

In this note we will be mostly
concerned with the phases of QCD at finite baryon density. Let me
then mention, as a last introductory remark, that
there are strong differences between
physical mechanisms of phase transitions in QCD at high baryon density and
high temperature. At high temperature the chiral transition is
a transition from an ordered to a disordered state, characterized
by light baryons; deconfinement is associated with string breaking
due to recombination with light pairs; topological structures
include instanton molecules.
At high baryon density instabilities
at the Fermi surface induce unusual patterns of chiral symmetry
such as color superconductivity
or superfluidity; string breaking might be further enhanced 
by recombination with real particles; topological structures 
are mostly likely  instanton chains.
These differences between  high temperature
and  high density  might provide further  insight into 
chiral symmetries,  
gluon dynamics and topological structures, and further motivate
the study of a nonzero baryon density on a lattice.

\section{Formulation}
Let us remind ourselves
  how to introduce a chemical 
potential $\mu$
for a conserved charge {$\hat N$} in the density matrix $\hat \rho$
in the Grand Canonical 
formalism, which is
the one appropriate for a relativistic field theory \cite{intro}:
\begin{eqnarray}
\hat \rho &=& e^{-(H - \mu \hat N)/T}   \\ 
\cal Z (T, \mu) & =& Tr \hat \rho = \int d \phi d \psi e^{-S(\phi, \psi)} 
\end{eqnarray}

The path integral representation of the grand partition function $\cal Z$  
in the Euclidean space gives the temperature as the reciprocal of the
imaginary time:
\begin{equation}
S(\phi, \psi) = \int_0^{1/T} dt \int d^d x {\cal L}(\phi, \psi) 
\end{equation}
with periodic boundary conditions in time for bosons 
$\phi(t=0,\vec x)  = \phi(t = 1/T, \vec x) $
and antiperiodic for fermions 
$ \psi(t=0,\vec x) = - \psi(t = 1/T, \vec x)$.

All in all, ${\cal Z}$ at finite temperature $T$  and density
$\mu$  is the partition function of a statistical system in d+1 dimension, 
where $T$ is the reciprocal of the imaginary time, and $\mu$ couples
to any conserved charge.
This representation, which is the starting point for a lattice
calculation,  allows us to deal with thermodynamics and spectrum
exactly on  the same footing.

The theory is regularised on a space time lattice: a regular 
four dimensional grid with $N_s$ points in each space directions,
$N_t$ points in the imaginary time direction, and spacing $a$.
 We refer to
the very many excellent reviews and textbooks for background material
on lattice field theory,  and
we briefly summarize here
the specific aspects of lattice QCD thermodynamics which will
be useful in the following.

The temperature $T$ on a lattice is the same as in 
the continuum: $T = 1 /N_ta$, $N_ta$ being the lattice extent
in the imaginary time direction (while, ideally, the lattice
spatial size should be infinite).
A lattice realisation 
of a finite density of baryons, instead, poses specific problems:
the naive discretization of the continuum expression 
$\mu \bar \psi \gamma_0 \psi$  
would  give an energy
$ \epsilon \propto \frac{\mu^2}{a^2}$ diverging 
in the continuum  (${a \to 0}$) limit \cite{form}.

The problem could be cured by introducing appropriate counterterms, however 
the analogy between $\mu$ and  an external field in the $0_{th}$ 
(temporal) direction offers a nicer solution by
considering the appropriate lattice conserved current \cite{form}. 
This amounts to the following
modification of the fermionic part of
the Lagrangian for the $0_{th}$direction  $L_F^0$:
\begin{equation}
 L_F^0(\mu) = \bar \psi_x \gamma_0 e^{\mu a}\psi_{x + \hat 0} -
      \bar \psi_{x + \hat 0} \gamma_0 e^{-\mu a}\psi_{x }
\end{equation}
while the remaining part of the Lagrangian is unchanged.
This yields the current:
\begin{equation}
J_0 = - \partial_\mu L =- \partial_\mu {L_F}^0(\mu) =
\bar \psi_x \gamma_0 e^{\mu a}\psi_{x + \hat 0} +
      \bar \psi_{x + \hat 0} \gamma_0 e^{-\mu a}\psi_{x } 
\end{equation}
This representation of $J_0$ is amenable  to a simple interpretation: the
time forward propagation is enhanced by $e^{\mu a}$, while
the time backward propagation is discouraged by $e^{-\mu a}$; hence, 
the link formulation generates a  particles--antiparticles asymmetry.
In addition,  note that  $\int J_0 = N - \bar N$ as it should.
An alternative way to look at the link formulation  introduces an
explicit dependence on the fugacity  $e^{\mu/T}$ 
via an unitary transformation for the fields 
\cite{Vink:1988vu}. 
In this way  $ L (\mu) = L (0) $, and the $\mu$ dependence is on 
the boundaries, via the fugacity $e^{\mu/T}$:
$\psi(x + N_T) = -e^{\mu a  N_T} \psi(x) = -e^{\mu/T} \psi(x)$.
This is analogous to the continuum case\cite{evans}.

\section{Calculational Schemes}
Having set up the formalism, the task is to  compute
\begin{equation}
{\cal Z}  = \int d U d \psi e^{-S(U, \psi)}
\end{equation}
where from now on the Lagrangian defining the Action will be that of
lattice QCD,
containing gluon fields $U$ and quark fields $\psi$.

We have two options. We might integrate out gluons first: 
\begin{equation}
\int dU d \psi d\bar \psi {\cal Z} (T, \mu, \bar \psi, \psi, U) 
\simeq \int  d \psi d\bar \psi {\cal Z} (T, \mu, \bar \psi, \psi) 
\end{equation}  
This produce an effective {approximate} fermion model:
the procedure is physically appealing, but not systematically improvable,
but for one special (lattice) case (see below).
Alternatively, 
we might integrate out fermions exactly, by taking advantage of the
bilinearity of the fermionic part of the Lagrangian
$L = L_{YM} + L_F = L_{YM} +  \bar \psi M (U) \psi$ :
\begin{equation} 
\int dU d \psi d \bar \psi {\cal Z} (T, \mu, \bar \psi, \psi, U)  = 
\int dU e ^{-(S_{YM}(U) - log(det M))} 
\end{equation}
The ``effective'' model we build this way is exact: the price to
pay being that its physical interpretation is not as clear as for effective
fermion models. Anyway, this expression is the starting point for numerical 
calculations: the fact that in many cases they are highly successful
tell us that the configuration space is well behaved enough that
only a minor subset of configurations, although carefully chosen via 
importance sampling,
suffice to produce reasonable results.

\subsection {Effective Fermionic Models: analytical approaches }

Let us start by following the first idea, namely integrating
out the gluon fields so to define an effective fermionic Action.
This is a time honored approach, leading, for instance, to 
the instanton model Hamiltonian, hence to the exciting
discoveries on the QCD phase diagram of the last five years\cite{effe} .

On the lattice, one very interesting approach leading
to a fermionic model is provided by
the strong coupling expansion: in the infinite gauge coupling limit 
the Yang Mills term decouples from the Action, and the
integral over the gauge fields can be carried out exactly.

 The starting point is the QCD lattice Lagrangian:
 \begin{eqnarray}
 S &=& -1/2 \sum_{x} \sum_{j=1}^3 
  \eta_j(x) [ \bar \chi(x) U_j(x) \chi(x + j) - \bar \chi (x + j)
 U^\dagger_j(x) \chi (x)]  \\
 && -1/2 \sum_x \eta_0(x) [ \bar \chi(x) U_0(x) \chi(x + 0) 
 - \bar \chi (x + 0) U^\dagger_0(x) \chi (x)]  \nonumber \\
 && -1/3 \sum_{x} 6/g^2 \sum_{\mu,\nu=1}^4[ 1 - re Tr U_{\mu \nu}(x)] 
 \nonumber \\
 && + \sum_x m \bar \chi \chi \nonumber
 \end{eqnarray}
 
 The $\chi, \bar \chi$ are the staggered fermion fields living on the
 lattice sites, the $U$'s are the $SU(N_c)$ gauge connections on the links, the
$\eta$'s are the lattice Kogut--Susskind
 counterparts of the Dirac matrices, and the chemical
 potential is introduced via the time link terms $e^\mu$, $e^{-\mu}$ 
as discussed above. This time
we have written down explicitly the lattice Action to show that the
 pure gauge term \\
 $ S_G = -1/3 \sum_{x} 6/g^2 \sum_{\mu,\nu=1^4 }[ 1 - re Tr U_{\mu \nu}(x)]$
 contains the gauge coupling in the denominator, hence it disappears in
 the infinite coupling limit. Consequently, one can perform independent
 spatial link integrations, leading to
 \begin{equation}
 {\cal Z} = \int \prod_{time links} dU_t 
 d \bar \chi d \chi e ^{-1/{4N} \sum_{ <x,y>}
 \bar \chi (x) \chi (x) \bar \chi (y) \chi (y) }
 e^{-S_t}
 \end{equation}
 where $\sum_{<x,y>}$ means sum over nearest neighboring links, terms
 of higher order have been dropped, and we recognize a four fermion
interaction \cite{strong}. Further manipulations yield 
the mean field  effective potential:
\begin{equation}
\nonumber V_{eff}(<\bar \psi \psi>, {\mu}) =  2 {cosh(rN_tN_c\mu)} + 
   sinh[(N_t+1)N_c <\bar \psi \psi>] 
/sinh(N_t < \bar \psi \psi>) 
\label{eq:sc}
\end{equation}
which we quote for further reference.
A standard  analysis of $V_{eff}$ finally gives the condensate as a function
of temperature and density, and allows the reconstruction of the
phase diagram. 

More recently this approach has been furthered both 
in two \cite{Nishida:2003uj}
and three colors\cite{Bringoltz:2002ug}, and
new developments on cluster algorithms have appeared as well 
\cite{Chandrasekharan:1999cm}. 

In order
to describe in detail the rich physics of the
finite density phase, one needs both to include
higher order terms into the strong coupling expansion, as well
as to go beyond a simple mean field analysis, which assumes an homogeneous
background.  
The question is as to whether such improved strong coupling approaches
would be able to generate a four fermion term with the correct
flavor structure as well as order of magnitude,
thus opening the possibility of a systematically improvable
approach to finite density QCD, including the study of the superconducting
phase.

\subsection{Effective Gluonic Models: Importance Sampling and the positivity
issue}

Let us write again
\begin{equation}
 {\cal Z}(T, \mu) = 
\int dU e ^{-(S_{YM}(U) - log(\det M))}
\end{equation}

When $\det M > 0$ the functional integral can be evaluated
with statistical methods, sampling the configurations according
to their importance $(S_{YM}(U) - log(\det M))$.
For this to be possible the would-be-measure  ($\det M $) has to be positive.

Let me mention at this point that the factorization method \cite{factor}
might alleviate the problems of complex measures by guiding
the simulations along a sensible path in the phase space.
I will not dwell on this interesting development which is not
really in the scope of an introductory review, but I wish to
call on it the attention of the interested reader, as it really
seems to offer some promise, and has been already tested in 
random matrix models.

In QCD with an even number of flavors, and zero chemical
potential,  standard importance sampling simulations are possible
if $\det M$ is real, which is true if
$M^\dagger = - P M P^{-1}$
where $P$ is any non singular matrix. 
In the most popular lattice fermion formulation this holds:
for Wilson fermions $P = \gamma_5$ and for staggered fermions  $P=I$
(note that this basically expresses a particle--antiparticle symmetry).
We will consider staggered fermions from now on.

Consider now the relationship 
$M^\dagger (\mu_B) = - M (-\mu_B) $
implying  that  reality is lost when $ {\rm Re} \mu \ne 0$:
the reality of the determinant is lost, and with it the possibility
of doing simulations with non zero chemical potential, when we want
to create a particle antiparticle asymmetry. On the other
hand a purely imaginary chemical potential does not spoil the
reality of the determinant: indeed, even if an imaginary chemical
potential can be used to extract information at real chemical
potential, it does not create any real particle--antiparticle
asymmetry and it is natural that the fermion determinant
remains real.

Note that in QCD with two color the determinant remains positive
with nonzero real chemical potential: indeed, in that case quarks and
antiquarks transform under equivalent representation of the
color group and are, essentially, the same particle. Other
important models with a real determinant include finite
density of isospin \cite{Sinclair:2003rm} and four fermion models
\cite{GNlattice}. These aspects will be reviewed
by Don Sinclair at this meeting \cite{Sinclair:2003rm}, 
so I shall not further discuss them.

All in all, if we want to extract information useful for QCD at
nonzero baryon density by use of standard MonteCarlo sampling
we will have to use information from the accessible region:
$$
 {\rm Re~~}   \mu =  0, {\rm Im~~}  \mu \le 0
$$

\section{Overview of the  methods}
To begin with, it is useful to think of the theory 
in the $T, \mu^2$ plane. Let us then discuss the phase diagram
from the perspective of analyticity and positivity of the partition
function and of the
determinant. One important consideration to keep in mind:
the Gran Canonical partition function has to 
be positive. It is only the determinant which can change
sign, or even be complex, on single configurations.

Let us consider a mapping 
from  complex $\mu$ to complex $\mu^2$.
Because of the symmetry properties of the theory, this mapping
can be done without loss of generality.
Let us note then that ${\cal Z}(\mu^2)$ is real valued for real  $\mu^2$:
this is a situation familiar from condensed matter: the partition function
is real where the external parameter is real, complex otherwise.

The reality region for the partition function represents states which
are physically accessible. The reality region for the determinant
represents the region which is amenable to an importance
sampling calculation: ${\rm Re}  \mu^2 \le 0$.
The methods which have been applied so far are
\begin{itemize}
\item {\bf $\mu=0$} Derivatives, Reweighting, Expanded reweighting 
\item {\bf $\mu^2 \le 0$} Imaginary chemical potential
\end{itemize}
\begin{wrapfigure}{r}{\halftext}
{\epsfig{file= 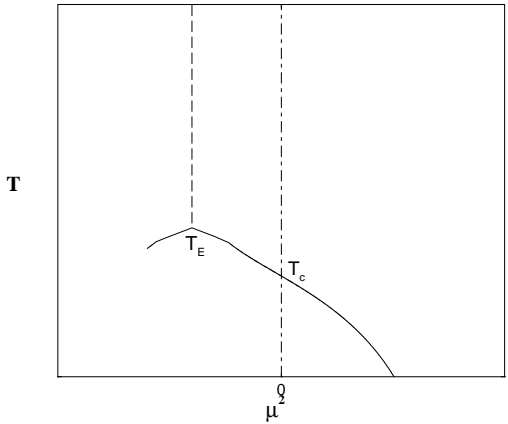, width= \halftext}}
\caption{Sketch of the phase diagram in the $\mu^2,T$ plane:
the solid line is the chiral transition, the dashed
line is the Roberge Weiss transition. Simulations can be carried out at
$\mu^2 \le 0$ and results continued to the
physical domain $\mu^2 \ge 0$.
The derivative and reweighting methods have been used so far
to extract informations from simulations performed at $\mu=0.$
The imaginary chemical potential approaches uses results on the
left hand half plane. Different methods could be combined 
to improve the overall performance.}
\end{wrapfigure}

\subsection{  Derivatives at $\mu=0.0$}
This is one early attempt at exploring the
physics of nonzero quark density: the derivatives
can be formally computed at $\mu=0$ \cite{deri}. 
The obvious limitation is that we do
not really know how far from the $\mu=0$ axis can we get.
Nonetheless,   such derivatives are interesting per se, and
the region where derivatives are clearly different from zero
is the natural candidate for the application of other
methods.

\subsection{ Reweighting from $\mu = 0$ }
Back in the 80's Ian Barbour and collaborators proposed
to calculate 
${\cal Z}(\mu)$ from  simulations at $\mu=0$:
\begin{equation}
{\cal Z} = \left<{{|M(\mu)|}\over {|M(\mu=0)|}}\right>_{\mu=0}
\end{equation}
In other words, the chemical potential $\mu$ of
the  target ensemble at that of the simulation ensemble 
-- $\mu=0$ --  are
different: the properties of the target ensemble can be inferred
from those of the simulation ensemble, provided that there is
a sizable overlap between the two \cite{Barbour:1997ej}.

At $T = 0$ the Glasgow procedure fails because of a poor overlap
(aside, the strong coupling calculations were quite useful to
asses these problems), and it is instructive to study
the overlap problem as seen in the Gross Neveu model, where there
is no sign problem \cite{GNlattice}, 
and the results obtained with reweighting methods
can be compared with those of exact simulations
\cite{Barbour:1999mc}.

\begin{figure}
{\epsfig{file= 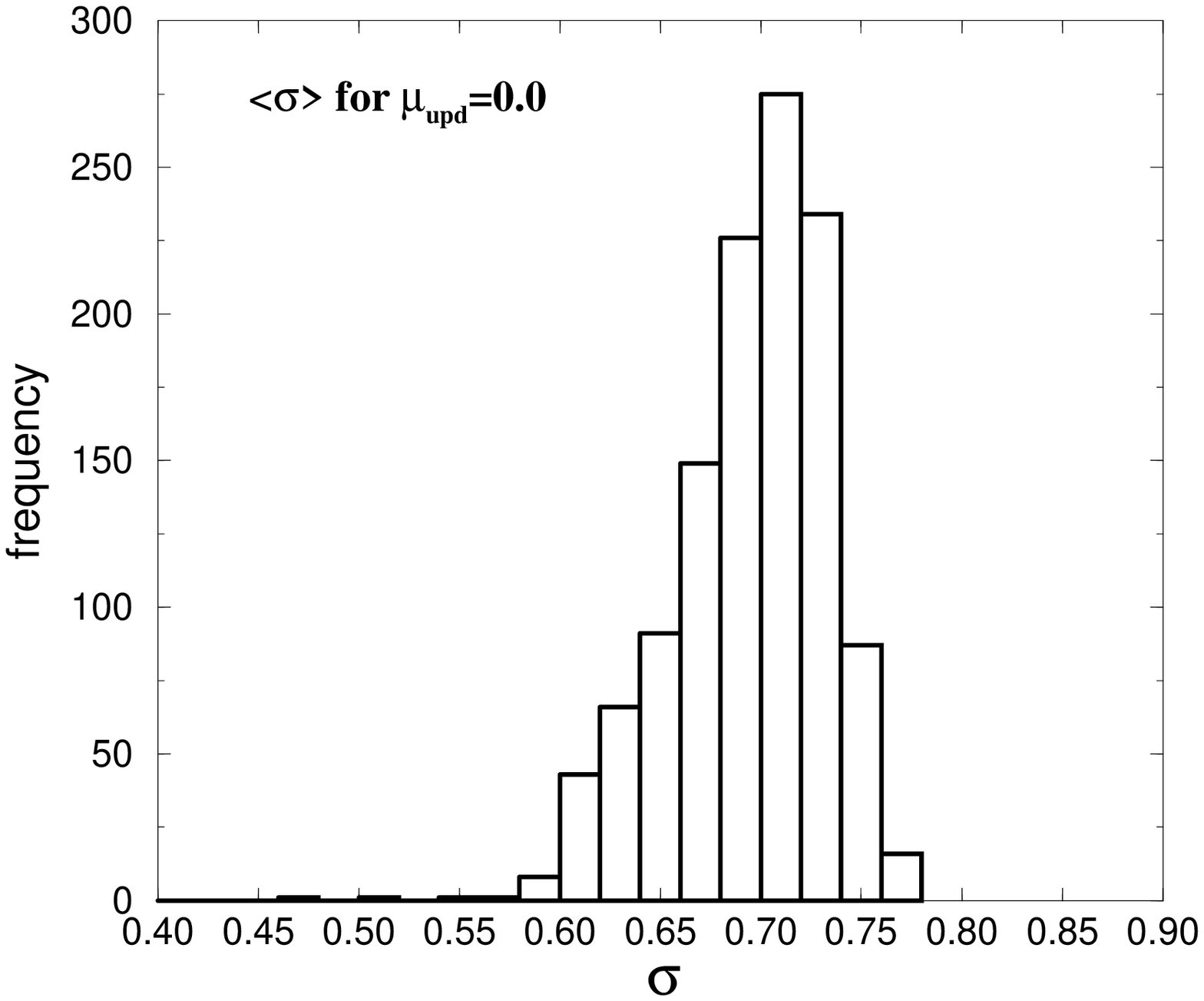, width= \halftext}}
{\epsfig{file= 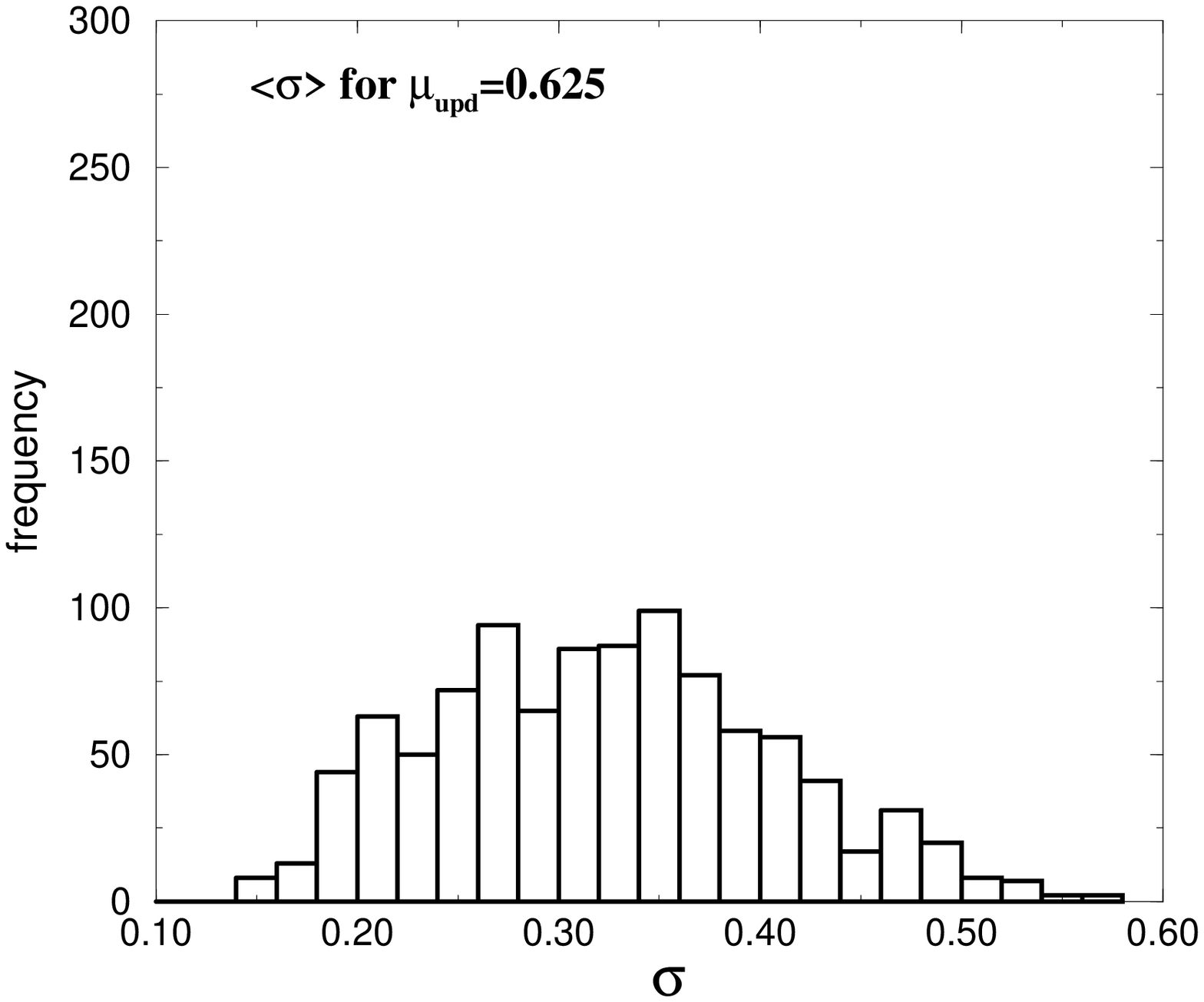, width= \halftext}}
\caption{The overlap problem in the Gross Neveu model\cite{Barbour:1999mc}: 
the exact
distributions of the $<\sigma>$ fields in the broken (left) and 
symmetric (right) phase are non--overlapping.} 
\label{fig:gnproblem}
\end{figure}

The distribution of the order parameter (the $\sigma$ particle)
helps visualizing the problem (see Fig. \ref{fig:gnproblem}): 
the order parameter distributions in the two phases do not 
overlap\cite{GNlattice}.
  
It is interesting to note that
 there are indeed examples of successful
reweighting at $\mu \ne 0$. Let us consider
1-dim $SU(3)$ which can be exactly solved at nonzero baryon density\cite{uno} 
and can
thus serve as a test bed for experiments. The distribution of the
partition function zeros was computed and found to reproduce 
the correct results one the statistics were sufficiently high\cite{Lombardo:1999cz}.
\begin{figure}
{\epsfig{file= 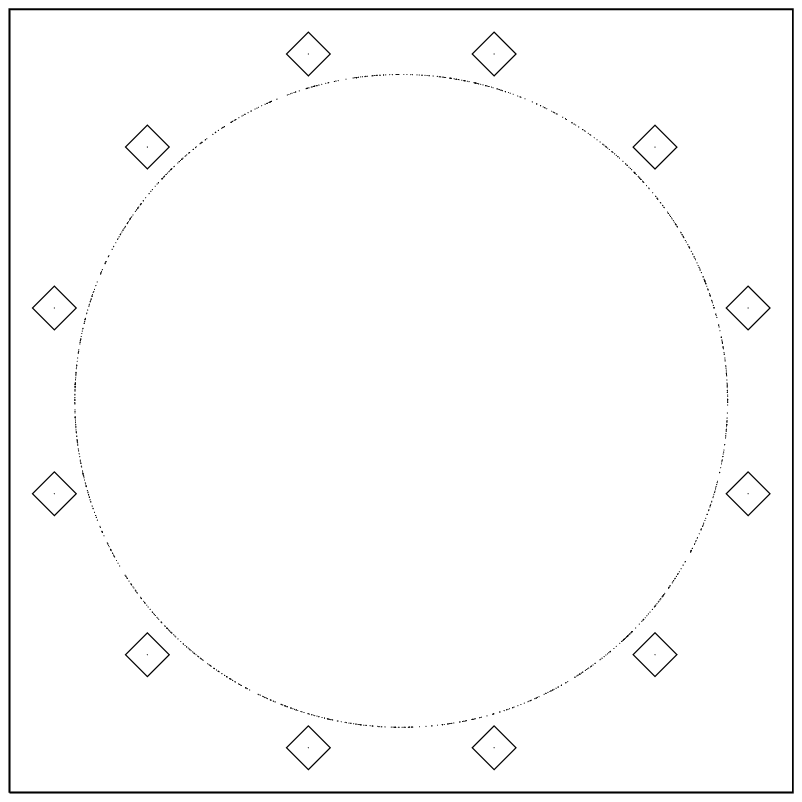, width= \halftext}}
{\epsfig{file= 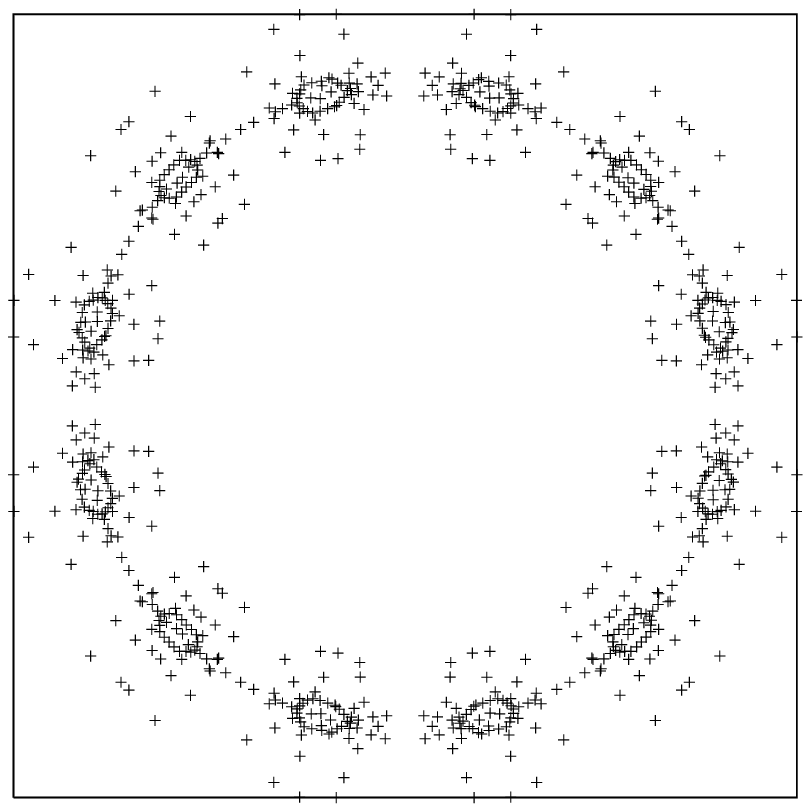, width= \halftext}}
\caption{Successful reweighting in one dimensional QCD\cite{Lombardo:1999cz}:
exact Z's zeros in 
the complex $\mu$ plane (diamonds) 
and the cloud of zeros obtained from reweighting with a very poor 
statistics(leftt). On the rigth 
the  zeros from an high statistics reweighting simulation
reproducing the exact result.}
\end{figure}

The conclusion from these early studies was that reweighting fails
in QCD at zero temperature because of a poor overlap, and that
the reason behind the failure is practical rather than conceptual:
the situation can be ameliorated if a better starting point were used.

\subsection{ Fodor and Katz's multiparameter reweighting }
The prescription for ameliorating the overlap is due to 
Fodor and Katz \cite{Fodor:2001au}\cite{Fodor-a}
whose {\em Multiparameter reweighting} 
use fluctuations around $T_c$ at $\mu=0$ to explore the critical region.
Making reference to Fig. 2, and oversimplifying: instead of trying
to reweight the distribution at zero temperature in the broken phase,
which is obviously hopeless, one might hope that a distribution
generated  at zero density, and close to the critical temperature, bears more
resemblance with the target distribution along the critical line,
and is thus amenable to a successful reweighting.

The strategy was applied to QCD \cite{Fodor:2001au}\cite{Fodor-a} .
The improvement obtained is impressive and produced the first
quantitative results for the critical line
at nonzero chemical potential in QCD: we will come back to this
in the section on results. A multistep reweighting proposed
by Crompton \cite{Crompton:2001ws}
might well produce a further improvement.

\subsection{Taylor  Expanded Reweighting}
The Bielefeld-Swansea collaboration suggested a
Taylor expansion of  the reweighting factor as a power series in 
$\lambda=\mu/T$,
and similarly for any operator\cite{Allton:2002zi}
\cite{Allton:2003vx} . 

This strategy is computationally very convenient
as it greatly simplifies the calculation of the determinant.
Expectation values are then given by
\begin{equation}
\langle{\cal O}\rangle_{(\beta,\mu)}=
{
{\langle({\cal O}_0+
 {\cal O}_1\lambda+{\cal O}_2\lambda^2+\ldots)
  \exp({\cal R}_1\lambda+{\cal R}_2\lambda^2+\ldots-\Delta S_g)\rangle_{\lambda=0,\beta_0}}
\over
{\langle\exp({\cal R}_1\lambda+{\cal R}_2\lambda^2+\ldots-\Delta S_g)
\rangle_{\lambda=0,\beta_0}}}.
\end{equation}
Results - to be discussed later- 
 have been obtained both for the critical line and thermodynamics.

\subsection{Imaginary Baryon Chemical Potential}

This method uses information from all of the negative $\mu^2$ half 
plane (Fig. 1) to explore the positive, physical relevant region.
An imaginary chemical potential $\nu$ in a sense bridges Canonical and
Grand Canonical ensemble\cite{cano}:
\begin{equation}
{\cal Z_C(N)} = \frac {\beta} {2 \pi} \int_0^{2 \pi/ \beta} d \nu {\cal Z_{GC}}
(i \nu) e^{- i \beta \nu N}
\end{equation}
The main physical idea behind any practical application is that 
at $\mu = 0$ fluctuations allow the exploration 
of $N_b \ne 0$ hence tell us about $\mu \ne 0$.
Mutatis mutandis, this is the same condition for the reweighting
methods to be effective: the physics of the simulation
ensemble has to overlap with that of the target ensemble.

A practical way to use the results obtained at negative
$\mu^2$ relies on their analytical continuation in the real
plane. For this to be effective\cite{Lombardo:1999cz}
${\cal Z} (\mu, T) $  must be analytical, nontrivial, and 
fulfilling this rule of thumb:
\begin{equation}
\chi(T,\mu) = \partial \rho (\mu, T) / \partial \mu 
= \partial^2 log Z (\mu, T) / \partial \mu^2
 > 0
\end{equation}

This approach has  been tested
in the strong coupling limit \cite{Lombardo:1999cz} of QCD, in the
dimensionally reduced model of high temperature QCD
\cite{Hart:2000ef}  and, more recently,   in the two
color model \cite{Giudice}.

Results (to be discussed later) have been obtained for
two \cite{deForcrand:2002ci}, four  \cite{D'Elia:2002gd}
and three \cite{deForcrand:2003hx} staggered flavors.

\section{Results }
The methods just outlined above are workarounds, not real
solutions: they are practical tools to circumvent a problem,
and, as such, it it not surprising that they  have to
be applied with a grain of salt, and that their performance
depends on the thermodynamic region which is being explored.

In this section I will go through the main issues
which have been addressed so far: the critical line,
the hadronic phase, the ``Roberge Weiss'' regime, the quark gluon
plasma phase, highlighting the main
strengths of the various methods alongside with the results.

\subsection{ The Critical Line}
The critical line has been obtained either by Fodor and Katz 
\cite{Fodor:2001au}\cite{Fodor-a} and by the 
Bielefeld Swansea collaboration within the multiparameter reweighting
or the expanded reweighting approach,
which gives  $Z(\mu)$ from $Z(0)$. The location of the end point
follows naturally within this framework, and its first determination
was given in \cite{Fodor-a} .

De Forcrand and Philipsen have also noticed that
the analytic continuation of the critical line from 
an imaginary $\mu$ is  possible \cite{deForcrand:2002ci},
and  have indicated and applied a strategy for the location of the
end point  \cite{deForcrand:2003hx}.

\begin{wrapfigure}{l}{\halftext}
{\epsfig{file= 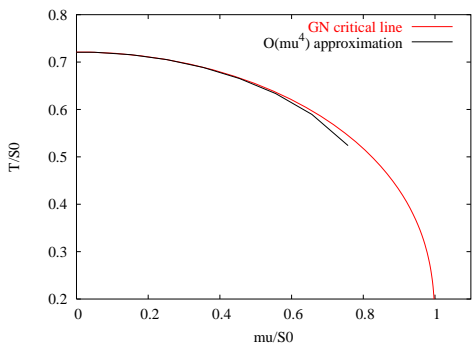, width= \halftext}}
\caption{The critical line of the 3d  Gross Neveu Model
and its polynomial approximation.}
\label{fig:gnfig}
\end{wrapfigure}

Also for the calculation of
the critical line the consideration 
of the $T,\mu^2$ plane helps the analysis \cite{D'Elia:2002gd}.
Model analysis suggests the following parametrization, confirmed by 
numerical results:
\begin{equation}
(T+aT_c)(T - T_c) + k \mu^2  = 0 ,\;\;\;\; k > 0
\end{equation}
It encodes reality for real $\mu^2$,
contains the physical scale $T_c$ , is dimensionally consistent, gives
$T(\mu = 0) = T_c$, $T(\mu \neq 0) < T_c$ .
For instance, the second order approximation $T(T - T_c) + \mu^2/(8ln2) = 0 $
to the Gross Neveu Model critical line: 
$1 - \mu /\Sigma_0 = 2 T /\Sigma_0 ln(1 + e^{-\mu/T}) $
is good up to $\mu \simeq T_c $, and from the plot in 
Fig. \ref{fig:gnfig}   can see that a
second order approximation is good up to $\mu \simeq T_c $

Studies of the critical line have indeed found that a simple
polynomial approximation suffices to describe the data, within
the current precision.  Progress on the precision is demonstrated
in Fig. 5 \cite{deForcrand:2003hx} 
where the new results on the three flavor model \cite{deForcrand:2003hx}
are superimposed to the older ones 
with two\cite{deForcrand:2002ci} and four flavors\cite{D'Elia:2002gd}.

A crucial issue remains the determination of the
endpoint, for which the first estimate was given within the
reweithgting method 
$T_E= 160 \pm 3.5 MeV$, $\mu_E = 725 \pm 35 MeV$ \cite{Fodor-a}.
Results with improved precisions show a dependence
on the mass values. This makes mandatory an extrapolation to physical
values of the quark masses, which , in turn, implies a good control
on the continuum limit.

\begin{wrapfigure}{l}{\halftext}
\centerline{
\includegraphics[width= \halftext]{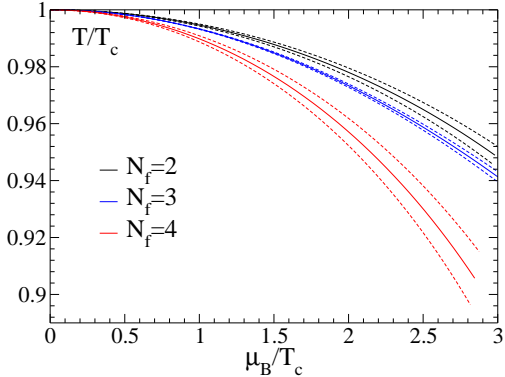}
} 
\caption{
Summary plot\cite{deForcrand:2003hx}  for the critical line for
$N_f=2$\cite{deForcrand:2002ci}, 
$N_f = 3$ \cite{deForcrand:2003hx}, 
$N_f = 4$ \cite{D'Elia:2002gd} from imaginary chemical potential
calculations. }
\label{fig:summary}
\end{wrapfigure}

\subsection{Hadronic Phase $T < T_c$ I: the hadron resonance gas model}

In this region observables are a continuous and periodic function
of $\mu_I/T$ , analytic continuation
in the $\mu^2 > 0$ half plane is always possible, but
interesting only when $\chi_q(\mu=0,T) > 0$.

Taylor expansion and Fourier decomposition are natural parametrization
for the observables\cite{D'Elia:2002gd}.
In particular, the  analysis of the phase diagram in the temperature-imaginary 
chemical potential plane suggests
to use Fourier analysis for $T \le T_c$, as observables are periodic
and continuous there. Note that in the infinite strong coupling
all of the Fourier coefficients but the first ones will be
zero (cfr eq. \ref{eq:sc}).

As the chiral condensate is an even function
of the chemical potential, its Fourier decomposition reads:
\begin{equation}
\langle \bar \psi \psi \rangle = \sum_n  a_F^{n} \cos (n N_t N_c \mu_I)
\end{equation}
which is easily continued to real chemical potential.

One cosine fit is actually enough to describe
the data up to $T \simeq T_c$ in the four flavor
model \cite{D'Elia:2002gd}: adding a
term  $\cos (2 N_t N_c \mu)$ in the expansion 
does not modify the value of
the first coefficients and does not particularly  improve the $\chi^2$.

This phase was also studied within the reweighting 
approach\cite{Karsch:2003zq}. It was confirmed that one single
hyperbolic cosine is an excellent approximation to the data,
and the result has been interpreted within the framework of the
hadron resonance gas model, whose  partition function 
has the single hyperbolic
cosine form as the one given by 
the strong coupling expansions eq. \ref {eq:sc}.

\subsection{The Hadronic Phase II: The order
of the phase transition, and related endpoints}

The analytic continuation of an observable $O$ is valid till $\mu < \mu_c(T)$,
where the critical value $\mu_c(T)$ has to be measured independently.
The value of the analytic continuation $O(\mu_c)$ 
of an observable $O$ at $\mu_c$
defines the discontinuity at the critical point.
In turns, this
allows the identification of the order of the phase transition.
One might  wonder which is
the meaning of the analytic continuation for $\mu > \mu_c(T)$, 
the one which we have to chop by hand.

It is  natural to interpret such analytic continuation
 as the metastable branch of
the observable we are considering, for instance $<\bar \psi \psi>$ : 
it  follows the secondary minimum of the associate Landau Ginsburg potential
and determines the spinodal point $\mu^*$ according to
$<\bar \psi \psi> = A(\mu - \mu^{*})^{\beta}$.
The discontinuity $<\bar \psi \psi>(\mu_c)$ 
is related to $(\mu_c - \mu^*)$, and
both shrinks to zero at the endpoint of a first order transition.

The analytic continuation of the results in the hadronic phase,
when cross examined with the results for the critical line,
offers an alternative way to study the order of the phase transition:
the transition will be second order is the zero of the analytic continuation
matches the critical point, first order otherwise. The endpoint of
a first order transition can be detected by monitoring 
$(\mu - \mu^{*})$ as a function of temperature.
\begin{wrapfigure}{r}{\halftext}
 \epsfig{file=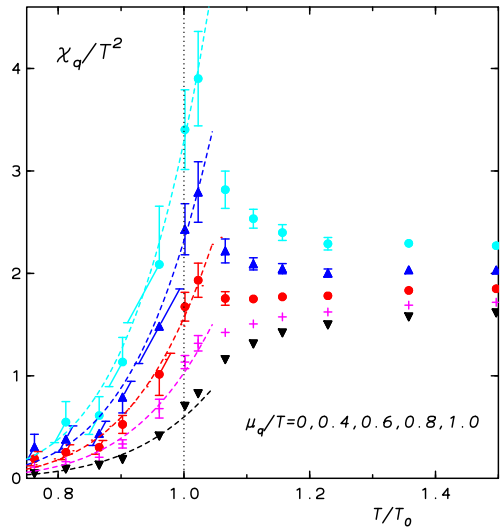,width= \halftext, angle=180}
\caption{The Bielefeld-Swansea
collaboration data
contrasted with the hadron resonance gas model \cite{Karsch:2003zq}.}
\end{wrapfigure}

Clearly this approach is specific to the imaginary chemical
potential calculations, and give results which should be cross
checked with others. For the time being it was confirmed 
\cite{D'Elia:2002gd} that the
transition in the four flavor model remains of first order at nonzero
density. 

\subsection{The Roberge Weiss Regime: $T_C < T < T_E$}
Let us consider region which is comprised between the deconfinement transition,
and the endpoint of the Roberge Weiss transition:
the analytic continuation is valid till $\mu = \infty$ 
but the interval accessible to the simulations for
$\mu^2 < 0 $ is small, as
simulations in this area hits the chiral critical
line.

The bright side of this is that the  nature of the critical line
can then be studied without need for an analytic continuation.
In Fig. 3 we show the clear correlation between the Polyakov Loop
and the chiral condensate 
at $\mu=0.15$\cite{D'Elia:2002gd}\cite{D'Elia:2003uy}.

\begin{wrapfigure}{r}{\halftext}
\psfig{file=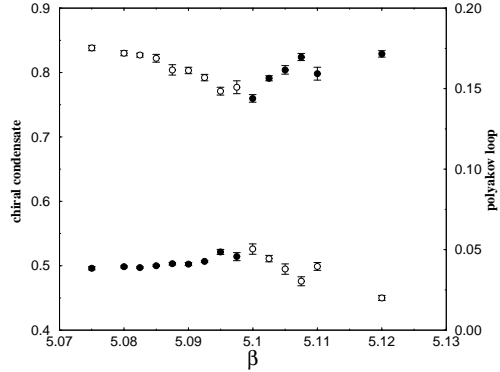,width=\halftext}
\caption{Correlation between $<\bar\psi\psi>$ and Polyakov loop
at $\mu_I = 0.15$ \cite{D'Elia:2003uy}. The observed 
correlation of chiral
and deconfining transition must persist
at nonzero baryon density\cite{D'Elia:2002gd}.}
\label{fig:polpsi}
\end{wrapfigure}

The correlation between
chiral and deconfining transition persists at nonzero imaginary chemical
potential, see Fig.  ~\ref{fig:polpsi}
\cite{D'Elia:2002gd}\cite{D'Elia:2003uy}. 
Obviously, these observations can be immediately
continued to real chemical potential : if the difference between the value
of the critical $\mu$'s for confinement and chiral symmetry is zero in a finite
interval within the analyticity domain (in our case, for $\mu^2 \le 0$), 
it will be zero everywhere within the same domain.

It is also of interest to note that the non--applicability of perturbation
theory in this region is almost a theorem: indeed the analytic continuation
of the polynomial predicted 
by perturbation theory for positive $\mu^2$ 
would never reproduce the correct critical behavior
at the second order phase transition for $\mu^2 < 0$ , and it
is then ruled out.

\subsection{The QGP phase : $T_E < T$}
At high temperature, in the weak coupling regime,
perturbation theory might serve as a guidance, 
suggesting that the first few terms of the
Taylor expansion might be adequate in
a wider range of chemical potentials.  
This confirms that the Roberge Weiss critical line
has to be strongly first order at high temperature.

Several analytic models have been proposed
to describe the properties of this 
phase.
As one example, we show how a
QGP liquid model prediction developed by 
Letessier and Rafelsky  \cite{Letessier:2003uj}:
\begin{equation}
\Delta P / T^4 = 2(1 - 2 \alpha_s/ \pi) ((\mu/T)^2)
\end{equation}
where $\alpha_s = \alpha_s(2 \sqrt{(\pi T)^2 + \mu_q^2})$,
compares favorably with reweighting data 
by Z.Fodor, S.Katz and K.K.Szabo \cite{Fodor:2002km}.
Very similar conclusions were reached by analyzing  quasi-particle
models \cite{quasi}.

As a further practical tool for the analysis of this phase, it was proposed
\cite{D'Elia:2003uy} to define an ``effective prefactor plot''
$ \Delta P K_{L(N_t=4)}/(T^4 2 (\mu/T)^2) $ 
 which should equal,
for instance  $2(1 - 2 \alpha_s/ \pi)$ in the model discussed above,
and, in general, which would serve 
to assess by eye  the $\mu$ dependence, if any, of the
prefactor of the quadratic term. 

It was found that the results 
approach the perturbative limit at large chemical potential, but corrections at
small chemical potential are clearly visible, and become much less
severe while approaching the large $\mu$ limit. It would be
interesting to understand the connection between these numerical observations
and the theoretical work on 
effective positivity on dense matter\cite{Hong:2002nn}.

\section{Summary/Outlook}

QCD at nonzero baryon density is now an active field
of research with results emerging from different methods.

The critical line for
two, three, two plus one, and four flavors has been computed
by various methods, with a substantial agreement.
The critical line is well described by a  polynomial,
and this result can be interpreted in terms of simple models. 

The endpoint has been located by different methods, and
its determination is currently being sharpened.
In the four flavor model,
when the transition is of first order, 
the chiral and deconfining transition remain correlated at
nonzero chemical potential.

The analytic continuation
from an imaginary chemical potential gives access to the
often evasive physics of the metastable branch. It 
might afford an alternative way to locate
the end point and tricritical point. 

Three different regimes have been considered and discussed:
the hadronic phase results are consistent with the hadron resonance gas model,
both from reweighting calculations and imaginary chemical potential approach;
the Roberge Weiss regime is eminently
nonperturbative, and in this regime we have the possibility to study
the nature of the chiral transition at nonzero chemical potential
without performing any analytic continuation; the
Quark Gluon Plasma phase seems well described by simple models,
but corrections are visible and still need being quantified.

Assessing the validity of simple models at high temperature/density,
besides being extremely interesting per se, would also open the
possibility of doing non equilibrium calculations based on these
models. The imaginary chemical potential approach seems to be particularly
well suited for this task.

\begin{wrapfigure}{r}{\halftext}
\epsfig{file=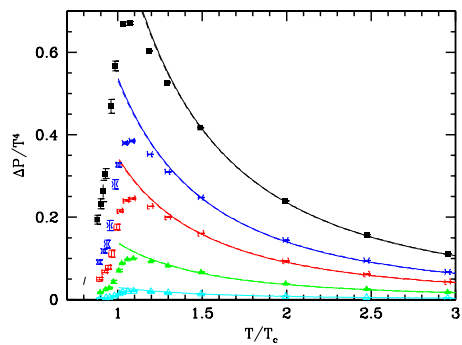,width=\halftext}
\caption{Lattice results by Fodor, Katz and Szabo contrasted with 
a Quark Gluon Plasma liquid model \cite{Letessier:2003uj}.}
\end{wrapfigure}
The three methods we have discussed
are mature for quantitative studies in realistic cases.
A nice possibility is offered by a combination of these methods:
 for instance either reweighting   or direct
calculations of derivatives could be performed 
at nonzero $\mu$ to improve the accuracy of the results at negative $\mu^2$,
and of the ensuing analytic continuation to real $\mu$.

Finally, it might well be that other methods 
such as $\chi QCD$ which in the past encountered
difficulties at zero temperature and nonzero baryon density 
will prove successful in the
richer high temperature regime \cite{Sinclair:2003rm}.

\section*{Acknowledgements}

It is a pleasure to thank the Organisers for this beautiful meeting
and their most kind hospitality in Nara.

\end{document}